\documentclass[a4paper,11pt]{article}
\pdfoutput=1 

\usepackage{jheppub} 

\usepackage[utf8]{inputenc}
\usepackage{bm}         
\usepackage{array}
\usepackage{multirow}   
\usepackage{booktabs}   
\usepackage{float}      
\usepackage{subcaption} 

\newcommand{\sgn}{\operatorname{sgn}}
\newcommand{\e}{\mathrm e}
\renewcommand{\d}{\mathrm d}

\renewcommand{\[}{\left[}

\newcommand{\Damu}{\Delta a_\mu}

\title{Muon g-2 and Dark Matter suggest Non-Universal Gaugino Masses: $\mathbf{SU(5)\times A_4}$ case study at the LHC}

\author[a,b]{Alexander S. Belyaev,}
\author[a]{Steve F. King,}
\author[a]{and Patrick B. Schaefers}

\affiliation[a]{School of Physics \& Astronomy, University of Southampton, Southampton SO17 1BJ, UK}
\affiliation[b]{Particle Physics Department, Rutherford Appleton Laboratory, Chilton, Didcot, Oxon OX11 0QX, UK}

\emailAdd{A.Belyaev@soton.ac.uk}
\emailAdd{S.F.King@soton.ac.uk}
\emailAdd{P.Schaefers@soton.ac.uk}

\abstract{
We argue that in order to account for the muon anomalous magnetic moment $g-2$, dark matter and LHC data, non-universal gaugino masses $M_i$ at the high scale are required in the framework of the Minimal Supersymmetric Standard Model (MSSM).
We also need a right-handed
smuon $\tilde\mu_R$ with a mass around 100 GeV,
evading LHC searches due to the
proximity of a neutralino $\tilde{\chi}^0_1$ several GeV lighter which allows successful dark matter.
We discuss such a scenario in the framework of an $SU(5)$ Grand Unified Theory (GUT)
combined with $A_4$ family symmetry, where the three $\overline{5}$ representations form a single triplet of $A_4$ with a unified soft mass $m_F$, while the three $10$ representations are singlets of $A_4$ with independent soft masses
$m_{T1}, m_{T2}, m_{T3}$.
Although $m_{T2}$ (and hence $\tilde\mu_R$) may be light, the muon $g-2$ and relic density also requires light $M_1\simeq 250$~GeV, which is incompatible with 
universal gaugino masses due to LHC constraints on $M_2$ and $M_3$ arising from gaugino searches.
After showing that universal gaugino masses $M_{1/2}$ at the GUT scale are excluded by gluino searches, 
we provide a series of benchmarks which show that while 
$M_{1}= M_{2} \ll M_3$ is also excluded by chargino searches, $M_{1}< M_{2} \ll M_3$ is currently allowed. 
Even this scenario is almost excluded by the tension
between the muon $g-2$, relic density, Dark Matter direct detection and LHC data.
The surviving parameter space is characterised by a higgsino mass $\mu \approx -300$ GeV,
as required by the muon $g-2$. The LHC will be able to fully test this scenario with the 
upgraded luminosity via muon-dominated tri- and di-lepton 
signatures resulting from higgsino dominated $\tilde{\chi}^\pm_1 \, \tilde{\chi}^0_2$ and 
$\tilde{\chi}^+_1 \, \tilde{\chi}^-_1$ production.
}


\begin{document} 
\maketitle
\flushbottom

\section{Introduction}
\label{sec:introduction}

The Minimal Supersymmetric Standard Model (MSSM) remains an attractive candidate for physics beyond the Standard Model (BSM) even in  the absence of any signal at the Large Hadron Collider (LHC). Despite the limits from direct and indirect searches for dark matter (DM), the lightest neutralino \cite{Jungman:1995df}, whose stability is enforced by R-parity, remains a prime candidate for the weakly interacting massive particle (WIMP). 

There are several constraints from the LHC that restrict the parameter space of the MSSM, in particular the requirement of a 125 GeV Higgs boson
and stringent limits on the gluino mass \cite{TheATLAScollaboration:2015cyl,Sirunyan:2017kqq}.

An interesting possible signature of BSM physics is the muon $g-2$ or anomalous magnetic moment $a_{\mu}=(g-2)_{\mu}/2$ which differs from its Standard Model (SM) prediction by amount \cite{Agashe:2014kda}:
\begin{eqnarray}
	\Delta a_{\mu}\equiv a_{\mu}({\rm exp})-a_{\mu}({\rm SM})= (28.8 \pm 8.0) \times 10^{-10}.
	\label{gg-22}
\end{eqnarray}
Although it is possible to account for the muon $g-2$ within a supersymmetric framework \cite{Belyaev:2016oxy,Grifols:1982vx,Ellis:1982by,Chakrabortty:2013voa,Chakrabortty:2015ika,Barbieri:1982aj,Kosower:1983yw,Yuan:1984ww,Romao:1984pn,Lopez:1993vi,Moroi:1995yh,Martin:2000cr,Czarnecki:2001pv,Baer:2004xx,Cho:2011rk,Endo:2011mc,Endo:2011xq,Endo:2011gy,Evans:2012hg,Endo:2013bba,Mohanty:2013soa,Ibe:2013oha,Akula:2013ioa,Okada:2013ija,Endo:2013lva,Bhattacharyya:2013xma,Gogoladze:2014cha,Kersten:2014xaa,Li:2014dna,Chiu:2014oma,Badziak:2014kea,Calibbi:2015kja,Kowalska:2015zja,Wang:2015rli}, it is well known that it cannot be achieved in the MSSM with universal soft masses consistent with the above requirements, and therefore, some degree of non-universality is required. For example, non-universal gaugino masses have been shown to lead to an acceptable muon $g-2$ \cite{Kawamura:2017amp,Akula:2013ioa}. 

In the framework of Grand Unified Theories (GUTs) such as $SU(5)$ and $SO(10)$, non-universal gaugino masses at $ M_{\rm GUT} $ can arise from non-singlet F-terms, or a linear combination of such terms \cite{Corsetti:2000yq,King:2007vh,Chattopadhyay:2009fr,Ananthanarayan:2007fj,Bhattacharya:2007dr,Martin:2009ad,Martin:2013aha,Anandakrishnan:2013cwa}. The most general situation is when all the gaugino masses may be considered as effectively independent. Recently, an $SU(5)$ model has been analysed with completely non-universal gaugino masses and two universal soft masses, namely $m_F$ and $m_T$, which accommodate the $\overline{5}$ and $10$ representations, respectively (with the two Higgs soft masses set equal to $m_F$) \cite{Ajaib:2017zba}. In such a framework it was shown that the muon $g-2$ and dark matter may both be explained successfully.

In this paper, we argue that in order to account for the muon anomalous magnetic moment and dark matter in supersymmetry, non-universal gaugino masses are required. In particular, $M_{1,2}\ll M_3$, even for non-universal scalar masses of the three families. In order to support this, we consider 
an $SU(5)$ Grand Unified Theory (GUT)
combined with an $A_4$ family symmetry, where the three $\overline{5}$ representations form a single triplet 
of $A_4$ with a unified soft mass $m_F$, while the three $10$ representations are singlets of $A_4$ with independent soft masses
$m_{T1}, m_{T2}, m_{T3}$. We show that, even with such family non-universality, it is not possible to account for the muon $g-2$
with universal gaugino masses. Allowing non-universal gaugino masses with $M_{1,2}\ll M_3$,
we show that, with $\mu \approx -300$ GeV,
it is possible to successfully explain
both the muon anomalous magnetic moment and dark matter, while remaining consistent with all other experimental constraints.
We present three benchmark points in our favoured region of parameter space involving a right-handed smuon mass around 100 GeV, which can decay into a bino-dominated neutralino plus a muon.
The remaining neutralino masses are all below about 300 GeV, while the rest of the SUSY spectrum has multi-TeV masses.

The layout of the remainder of the paper is as follows. In section~\ref{sec:the-model}, we present the $SU(5) \times A_4$ model and its symmetry breaking to the MSSM. In section~\ref{sec:g-2}, we summarise the MSSM one-loop contributions to $\Damu$ and give first predictions for viable regions of parameter space of the model. All experimental constraints we take into account (both collider and cosmological constraints) are listed and explained in section~\ref{sec:constraints}. In section~\ref{sec:results}, we present scans of the model parameter space for universal and non-universal gaugino masses, which also helps clarifying the necessity of non-universal gaugino masses. Lastly, we draw our conclusions in section~\ref{sec:conclusions}.

\section{The Model}
\label{sec:the-model}

We first consider the gauge group $SU(5)$,
which is rank 4 and has 24 gauge bosons which transform as the ${\bf 24}$ adjoint representation.
A LH lepton and quark fermion family is neatly accommodated into the 
$SU(5)$ representations 
$F=\overline{\bf 5}$ and $T={\bf 10}$, where 
\begin{equation}
F= \left(
\begin{array}{c}d_r^c\\d_b^c\\d_g^c\\e^-\\-\nu_e \end{array}\right)_L  , \qquad
T= \left(
\begin{array}{ccccc} 0&u_g^c&-u_b^c&u_r&d_r\\
.&0&u_r^c&u_b&d_b\\ 
.&.&0&u_g&d_g\\ 
.&.&.&0&e^c\\
.&.&.&.&0
\end{array}\right)_L  \ ,
\end{equation}
where $r,b,g$ are quark colours and 
$c$ denotes CP conjugated fermions. 

The $SU(5)$ gauge group may be broken to the SM 
by a Higgs multiplet in the ${\bf 24}$ representation developing a VEV,
\begin{equation}
SU(5)\rightarrow SU(3)_C\times SU(2)_L\times U(1)_Y ,
\end{equation}
with
\begin{equation}
\overline{\bf 5}=d^c(\overline{\bf 3},{\bf 1},1/3)\oplus  L({\bf 1},\overline{\bf 2},-1/2),
\end{equation}
\begin{equation}
{\bf 10} =u^c(\overline{\bf 3},{\bf 1},-2/3)\oplus  
Q({\bf 3},{\bf 2},1/6)\oplus e^c({\bf 1},{\bf 1},1),
\end{equation}
where $(Q,u^c,d^c,L,e^c)$ is a complete quark and lepton SM family.
Higgs doublets $H_u$ and $H_d$, 
which break EW symmetry in a two Higgs doublet model, may arise from
$SU(5)$ multiplets $H_{\bf 5}$ and $H_{\overline{\bf 5}}$,
providing the colour triplet components can be made heavy.
This is known as the doublet-triplet splitting problem.

When $A_4$ family symmetry is combined with $SU(5)$, it is quite common to unify the three 
families of $\bar{5} \equiv F \equiv   (d^c, L)$ into a triplet of $A_4$, 
with a unified soft mass $m_F$, while the three $10_i \equiv T_i \equiv (Q, u^c, e^c)_i$ representations are singlets of $A_4$ with independent soft masses $m_{T1}, m_{T2}, m_{T3}$ \cite{Callen:2012kd,Antusch:2013wn,Cooper:2012wf,Cooper:2010ik,Bjorkeroth:2015ora}.
For simplicity, we will assume that at the GUT  scale  we have  $m_F = m_{H_u} = m_{H_d}$, where  $m_{H_u}$ and $m_{H_d}$ are the mass parameters of the MSSM Higgs doublets.

In the considered $SU(5)\times A_4$ model we then have the soft scalar masses:
\begin{eqnarray}
 &&  m_F = m_{\tilde{D}^c_i} =  m_{\tilde{L_i}} = m_{H_u} = m_{H_d} \,,
\nonumber \\
 &&  m_{T1}=m_{\tilde{Q_1}} = m_{\tilde{U_1}^c} = m_{\tilde{E_1}^c} \,,
 \nonumber \\
 &&  m_{T2}=m_{\tilde{Q_2}} = m_{\tilde{U_2}^c} = m_{\tilde{E_2}^c} \,,
 \nonumber \\
 &&  m_{T3}=m_{\tilde{Q_3}} = m_{\tilde{U_3}^c} = m_{\tilde{E_3}^c} \,.
\label{SU5A4}
\end{eqnarray}
Notice that the stop mass parameters are completely contained in $m_{T3}$, while the right-handed smuon mass arises from
$m_{T2}$, and so on.

\section{\texorpdfstring{MSSM One-loop contributions to $\Damu$}{MSSM One-loop contributions to Delta a\_mu}}
\label{sec:g-2}
\begin{figure}[htbp!]
	\centering
	\begin{subfigure}[t]{0.49\textwidth}
		\centering
		\includegraphics[width=\linewidth]{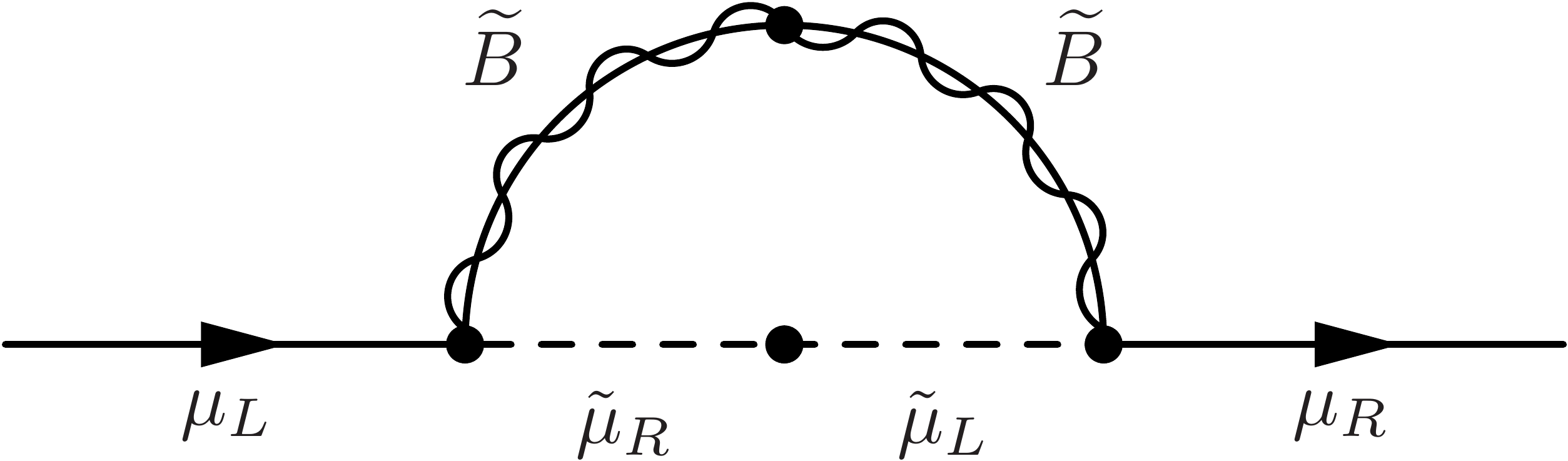}
		\caption{}
	\end{subfigure}
	\hfill
	\begin{subfigure}[t]{0.49\textwidth}
		\centering
		\includegraphics[width=\linewidth]{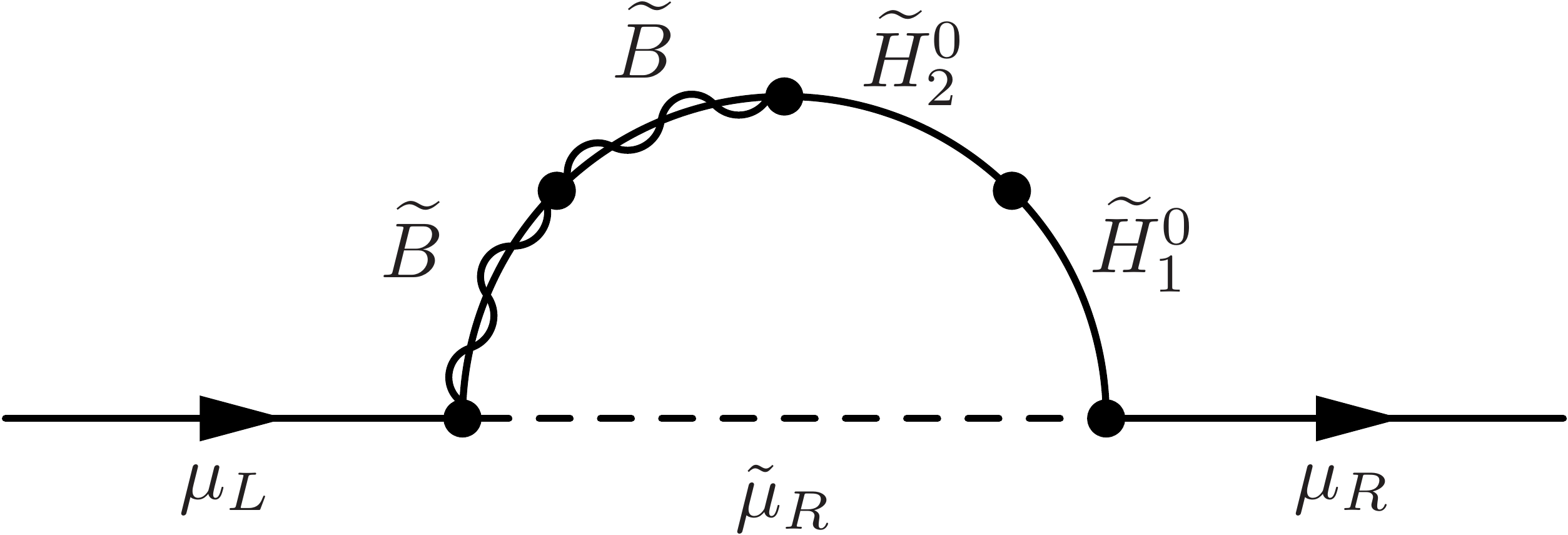}
		\caption{}
	\end{subfigure}
	\\[.3cm]
	\begin{subfigure}[t]{0.49\textwidth}
		\centering
		\includegraphics[width=\linewidth]{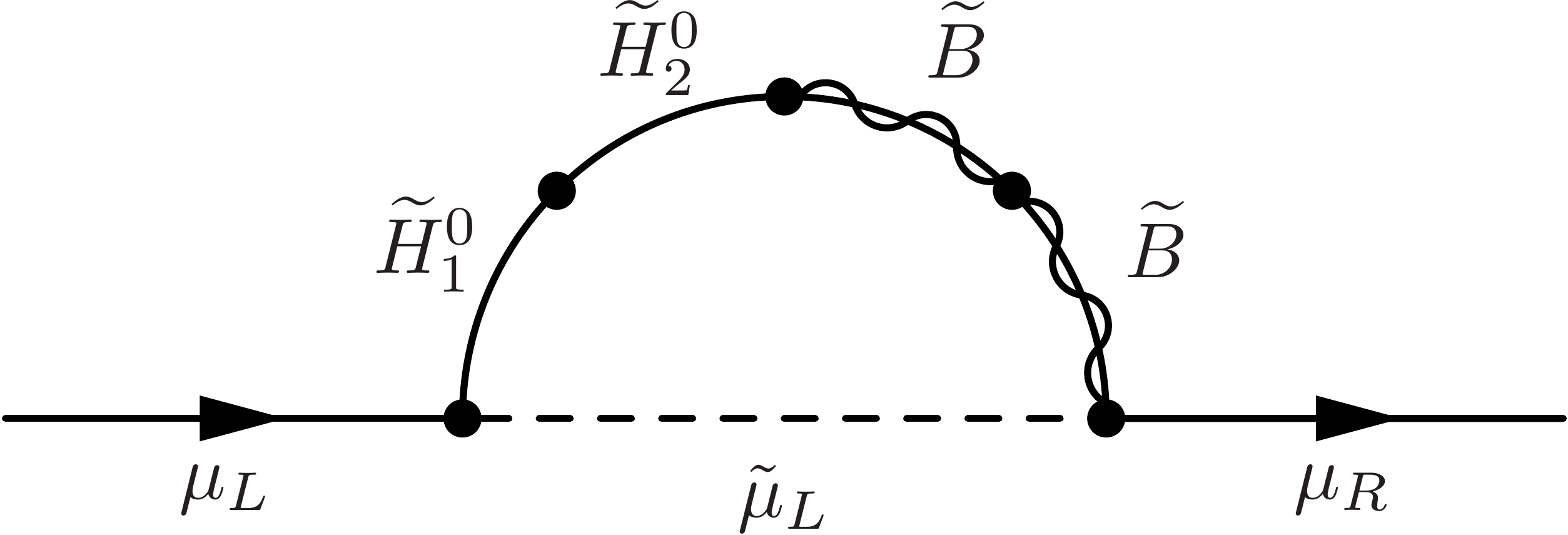}
		\caption{}
	\end{subfigure}
	\hfill
	\begin{subfigure}[t]{0.49\textwidth}
		\centering
		\includegraphics[width=\linewidth]{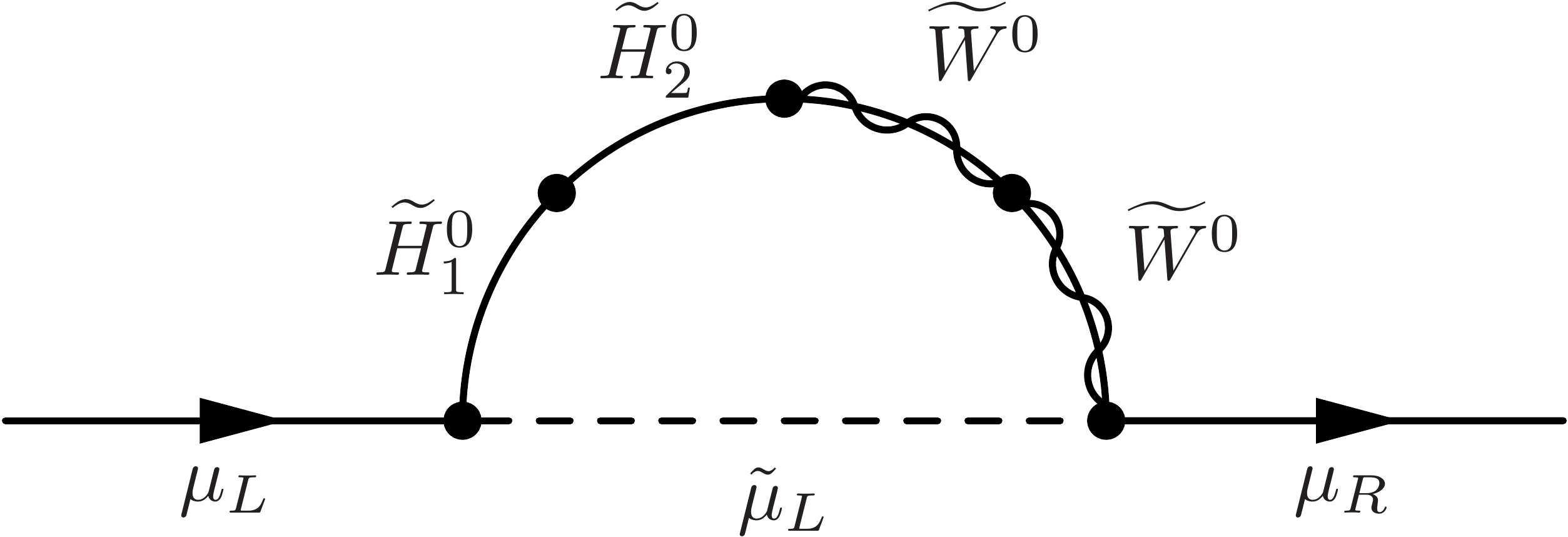}
		\caption{}
	\end{subfigure}
	\\[.3cm]
	\begin{subfigure}[t]{0.49\textwidth}
		\centering
		\includegraphics[width=\linewidth]{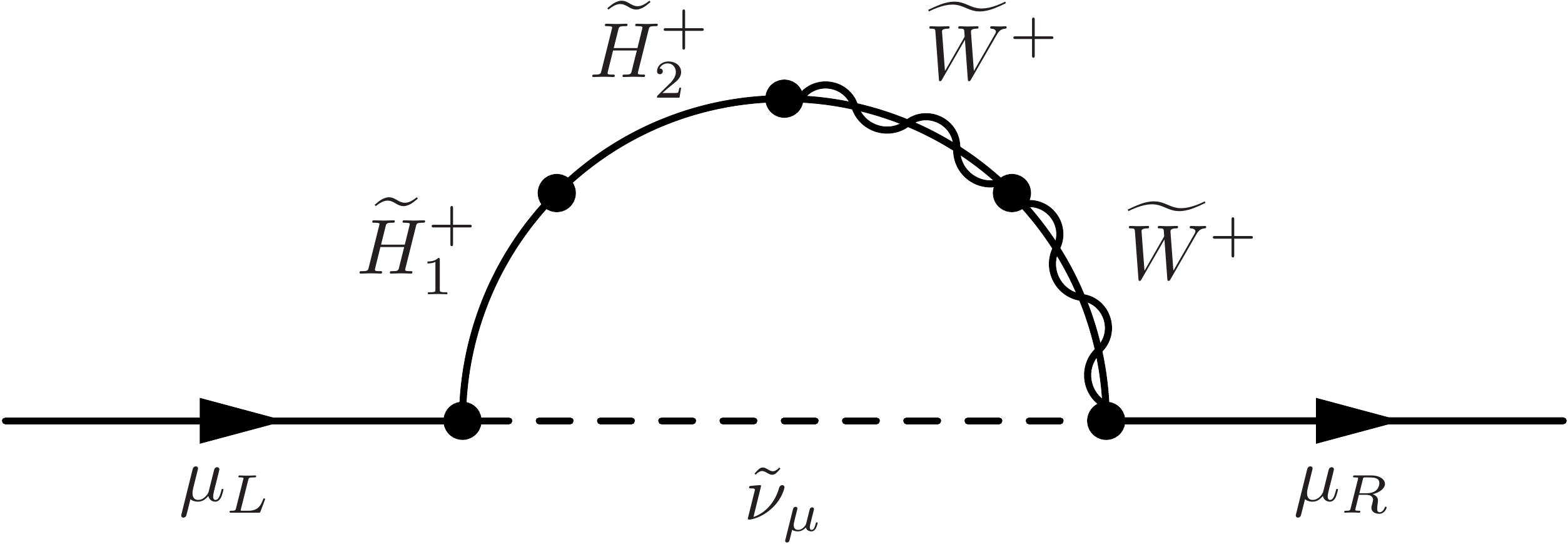}
		\caption{}
	\end{subfigure}
	\caption{One-loop contributions to the anomalous magnetic moment of the muon for supersymmetric models with low-scale MSSM.}
	\label{fig:1-loop-damu}
\end{figure}

The Feynman diagrams for the one-loop contributions to $\Damu$ in the MSSM are shown in figure~\ref{fig:1-loop-damu} with the respective expression for each diagram given by equations~\ref{eq:loops:A}~--~\ref{eq:loops:E} \cite{Moroi:1995yh,Endo:2013bba}. 

\begin{subequations}
\label{eq:loops}
\begin{align}
\Delta a_{\mu}^{(A)}&=\left(\dfrac{M_{1}\mu}{m_{\tilde{\mu}_{L}}^{2}m_{\tilde{\mu}_{R}}^{2}}\right)\dfrac{\alpha_{1}}{4\pi}m_{\mu}^{2}\tan\beta\cdot f^{(A)}_{{\rm N}}\left(\dfrac{m_{\tilde{\mu}_{L}}^{2}}{M_{1}^{2}},\dfrac{m_{\tilde{\mu}_{R}}^{2}}{M_{1}^{2}}\right)\, , \label{eq:loops:A} \\  
\Delta a_{\mu}^{(B)}&=-\left(\dfrac{1}{M_{1}\mu}\right)\dfrac{\alpha_{1}}{4\pi}m_{\mu}^{2}\tan\beta\cdot f^{(B)}_{{\rm N}}\left(\dfrac{M_{1}^{2}}{m_{\tilde{\mu}_{R}}^{2}},\dfrac{\mu^{2}}{m_{\tilde{\mu}_{R}}^{2}}\right)\, , \label{eq:loops:B} \\
\Delta a_{\mu}^{(C)}&=\left(\dfrac{1}{M_{1}\mu}\right)\dfrac{\alpha_{1}}{8\pi}m_{\mu}^{2}\tan\beta\cdot f^{(C)}_{{\rm N}}\left(\dfrac{M_{1}^{2}}{m_{\tilde{\mu}_{L}}^{2}},\dfrac{\mu^{2}}{m_{\tilde{\mu}_{L}}^{2}}\right)\, , \label{eq:loops:C} \\
\Delta a_{\mu}^{(D)}&=-\left(\dfrac{1}{M_{2}\mu}\right)\dfrac{\alpha_{2}}{8\pi}m_{\mu}^{2}\tan\beta\cdot f^{(D)}_{{\rm N}}\left(\dfrac{M_{2}^{2}}{m_{\tilde{\mu}_{L}}^{2}},\dfrac{\mu^{2}}{m_{\tilde{\mu}_{L}}^{2}}\right)\, , \label{eq:loops:D} \\
\Delta a_{\mu}^{(E)}&=\left(\dfrac{1}{M_{2}\mu}\right)\dfrac{\alpha_{2}}{4\pi}m_{\mu}^{2}\tan\beta\cdot f^{(E)}_{{\rm C}}\left(\dfrac{M_{2}^{2}}{m_{\tilde{\nu}_{\mu}}^{2}},\dfrac{\mu^{2}}{m_{\tilde{\nu}_{\mu}}^{2}}\right)\, . \label{eq:loops:E}
\end{align}
\end{subequations}
Here, $\alpha_1$ and $\alpha_2$ label the $U(1)_{Y}$ and $SU(2)_L$ fine structure constants respectively and the functions $f^{(A,B,C,D)}_{{\rm N}}\left(x,y\right)$ and $f^{(E)}_{{\rm C}}\left(x,y\right)$ are given by
\begin{subequations}
\label{eq:f}
\begin{align}
f^{(A,B,C,D)}_{{\rm N}}\left(x,y\right)&=xy\left[\dfrac{-3+x+y+xy}{\left(x-1\right)^{2}\left(y-1\right)^{2}}+\dfrac{2x\log x}{\left(x-y\right)\left(x-1\right)^{3}}-\dfrac{2y\log y}{\left(x-y\right)\left(y-1\right)^{3}}\right]\, , \label{eq:f:1} \\
f^{(E)}_{{\rm C}}\left(x,y\right)&=xy\left[\dfrac{5-3\left(x+y\right)+xy}{\left(x-1\right)^{2}\left(y-1\right)^{2}}-\dfrac{2\log x}{\left(x-y\right)\left(x-1\right)^{3}}+\dfrac{2\log y}{\left(x-y\right)\left(y-1\right)^{3}}\right]\, , \label{eq:f:2}
\end{align}
\end{subequations}
where we use the superscripts $(A,B,C,D)$ and $(E)$ as a short notation to allow omission of the mass ratio arguments. Both $f_N$ and $f_C$ are monotonically increasing for all $0 \leq (x,y) < \infty$ and are defined in $0 \leq f_{N,C} \leq 1$ \cite{Endo:2013bba}. 

One of the most important parameters influencing $\Damu$ is $\mu$, or rather $\sgn \mu$. Having positive $\mu$ means positive contributions from diagrams (A), (C) and (E), whereas negative $\mu$ results in (B) and (E) giving positive contributions to $\Damu$. Although it has been shown in the past that the constrained MSSM (cMSSM) with its usual five parameters ($M_{1/2}, m_0, \tan \beta, A_0, \sgn \mu$) is able to yield the observed $\Damu$, it cannot account simultaneously for further experimental limits (see e.g. \cite{Moroi:1995yh,Ibe:2013oha,Endo:2013bba}), regardless of $\sgn \mu$ but especially not for negative $\mu$. Extending the cMSSM or relaxing some of its constraints changes the picture and new solutions without the need for fine tuning arise --- all while being in conformity with all other low energy observations \cite{Belyaev:2016oxy,Grifols:1982vx,Ellis:1982by,Chakrabortty:2013voa,Chakrabortty:2015ika,Barbieri:1982aj,Kosower:1983yw,Yuan:1984ww,Romao:1984pn,Lopez:1993vi,Moroi:1995yh,Martin:2000cr,Czarnecki:2001pv,Cho:2011rk,Endo:2011mc,Endo:2011xq,Endo:2011gy,Evans:2012hg,Endo:2013bba,Mohanty:2013soa,Ibe:2013oha,Akula:2013ioa,Okada:2013ija,Endo:2013lva,Bhattacharyya:2013xma,Gogoladze:2014cha,Kersten:2014xaa,Li:2014dna,Chiu:2014oma,Badziak:2014kea,Calibbi:2015kja,Kowalska:2015zja,Wang:2015rli}. 

In this work, we have found that only the negative $\mu$ solution survives.
The reason why only negative $\mu$ survives is because in this case,
we are able to have light right-handed smuons while the left-handed smuons remain
rather heavy. This means that we are able to enhance the contribution from diagram (B) in which the right-handed smuons (but not the left-handed
smuons) appear. As already mentioned, negative $\mu$ results in diagram (B) giving a positive contribution to $\Damu$ and this is the main reason
why we favour negative $\mu$. 
In general, for negative $\mu$, the contribution from  diagrams (B) and (D) is enhanced, while  all contributions from diagrams (A), (C) and (E) (see section~\ref{sec:g-2})  are simultaneously suppressed. Enhancing (B) and (D) requires small $|\mu|$ (not directly controllable), small $M_{1}$ and $M_2$ as well as light left- and right-handed smuon masses $m_{\tilde{\mu}_L}$ and $m_{\tilde{\mu}_R}$ (controlled by $m_F$ or $m_{T2}$ respectively). On the other hand, light $m_{\tilde{\mu}_L}$ would lead to unwanted large contributions from diagrams (A) and (C). This is one reason to not have light $m_{\tilde{\mu}_L}$, but make them rather heavy. Another reason for heavy $m_{\tilde{\mu}_L}$ comes from the model parameter space itself. Since $m_{\tilde{\mu}_L}$ is governed by $m_{F}$, which also controls the muon sneutrino mass $m_{\tilde{\nu}^\mu_L}$ appearing in diagram (E), it is possible to decrease contributions from diagrams (A), (C) and (E) in one go by setting $m_{F}$ large. 

In the next section we briefly summarise the experimental constraints,
before discussing the full results in detail in section~\ref{sec:results}.

\section{Experimental Constraints}
\label{sec:constraints}
While the underlying model is proposed for the high-energy sector, it should nevertheless comprise any low-energy observations and limits coming from various experiments. In particular, we take into account the Dark Matter relic density, Dark Matter direct detetion (DD) cross sections, the Higgs boson mass, constraints coming from Br$(B_S \to \mu^+ \mu^-)$ as well as Br$(b \to s \gamma)$ and several 8 and 13 TeV \textsc{ATLAS} and \textsc{CMS} searches at the LHC.
Regarding the DM relic density, the current combined best fit to data from \textsc{PLANCK} and \textsc{WMAP} is $\Omega h^2 = 0.1198 \pm 0.0026$ \cite{Ade:2013zuv} and we consider a parameter space with $\Omega h^2\lesssim 0.1224$.

The current best DM DD limit comes from the \textsc{XENON1T} experiment, reading $\sigma_{\text{DD-SI}} \leq 7.64 \times 10^{-47}\ \text{cm}^2 = 7.64 \times 10^{-11} \ \text{pb}$ \cite{Aprile:2017iyp} for spin-independent models and a WIMP-mass of 36 GeV. Since WIMP masses smaller or larger than 36 GeV lead to weaker limits, this choice is conservative.
Concerning the Higgs boson mass, the current combined \textsc{ATLAS} and \textsc{CMS} measurement is $m_h = (125.09 \pm 0.21\, (\text{stat.}) \pm 0.11\, (\text{sys.})) \ \text{GeV}$ \cite{Aad:2015zhl}. However, due to the theoretical error in the radiative corrections to the Higgs mass inherent
in the existing state of the art SUSY spectrum generators, we consider instead the larger range 
$m_h = (125.09 \pm 1.5)$ GeV, which encompasses the much larger theoretical uncertainties.
The branching ratios $\text{Br}(b \to s \gamma) = (3.29 \pm 0.19 \pm 0.48) \times 10^{-4}$ \cite{Lees:2012wg} and $\text{Br}(B_s \to \mu^+ \mu^-)$$ = 3.0^{+1.0}_{-0.9} \times 10^{-9}$ \cite{Chatrchyan:2013bka} are directly applied to our results.
     
\section{Results}
\label{sec:results}

Following the strategy to enhance $\Damu$
in section~\ref{sec:g-2} and the experimental constraints in section~\ref{sec:results},
we are left with the following desired choice of parameters:
\begin{itemize}
	\item $m_{F}$ large for large $m_{\tilde{\mu}_L}$ and $m_{\tilde{\nu}^\mu_L}$,
	\item $m_{T2}$ small for light $m_{\tilde{\mu}_R}$,
	\item $m_{T1}$ and $m_{T3}$ large for large squark masses,
	\item $M_{1}$ small for light $\tilde{\chi}^0_1$,
	\item $\tan \beta$ large (affects all diagrams).
\end{itemize}
All other parameters are in principle unconstrained, but in practice will be constrained by experiment.

To gather the data for this work, we used \texttt{SPheno\_v4.0.3} \cite{Porod:2003um,Porod:2011nf} to generate the mass spectra based on input points chosen randomly as well as on fixed grids with variable grid spacing in the parameter space from tables~\ref{tab:univ-scan} and \ref{tab:nonuniv-scan} below. Subsequently, we employ \texttt{micrOMEGAs\_v3.6.9.2} \cite{Belanger:2013oya} to compute $\Damu$ and the low-energy constraints listed in section~\ref{sec:constraints}.
In the following two subsections, we present scans taking these considerations into account. Subsection~\ref{sec:results-universal} holds data and results regarding fully universal gaugino masses, commonly labelled as $M_{1/2}$, whereas subsection~\ref{sec:results-nonuniversal} refers to the case of partially non-universal gaugino masses, labelled as $M_{1,2}$ and $M_3$, and subsection~\ref{sec:results-fully-nonuniversal} refers to the case of fully non-universal gaugino masses labelled $M_1$, $M_2$ and $M_3$.

\subsection{Universal Gaugino Masses}
\label{sec:results-universal}
\begin{figure}
	\centering
	\includegraphics[width=0.49\linewidth]{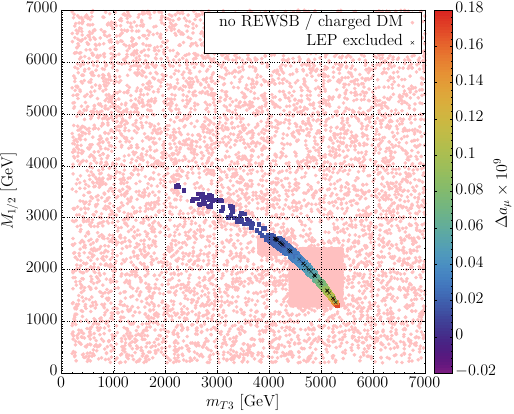}
	\hfill
	\includegraphics[width=0.49\linewidth]{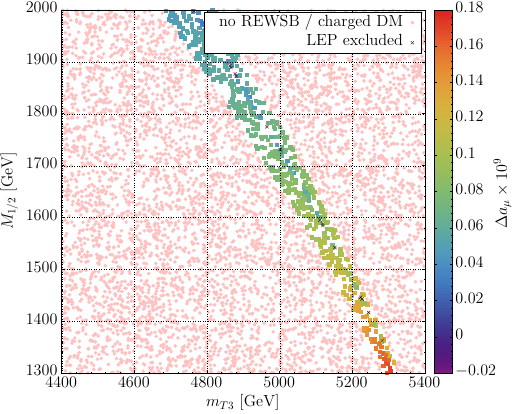}
	\caption{$m_{T3}$-$M_{1/2}$ plane with colour-coded $\Damu$ with universal gaugino masses. The right panel is an excerpt of the full scan shown in the left panel.}
	\label{fig:m3_vs_M3_vs_gmuon-univ}
\end{figure}

\begin{figure}
	\centering
	\includegraphics[width=0.49\linewidth]{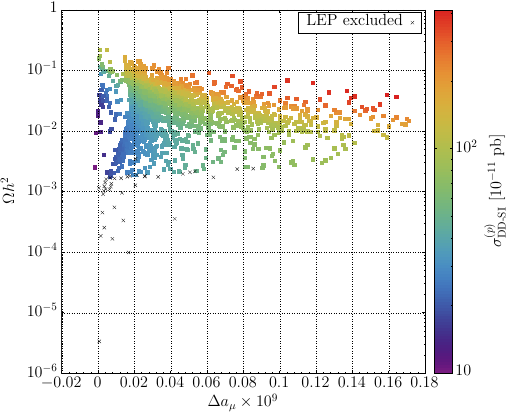}
	\hfill
	\includegraphics[width=0.49\linewidth]{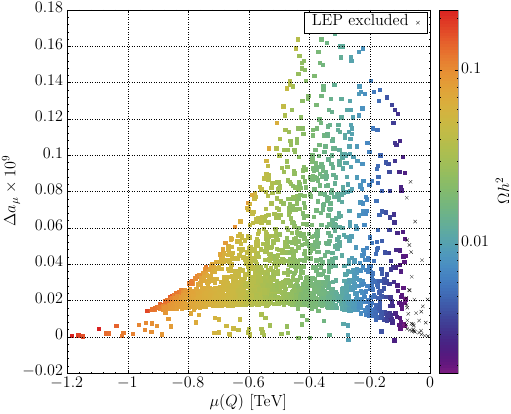}
	\caption{Left: Relic density vs. $\Damu$ with colour-coded $\sigma^{(p)}_{\text{DD-SI}}$ with universal gaugino masses. Right: $\Damu$ vs. $\mu$ with colour-coded relic density $\Omega h^2$ with universal gaugino masses.}
	\label{fig:gmuon_vs_Omega_univ}
\end{figure}

The scan with universal gaugino masses $M_{1/2}$ was performed with 
\begin{align*}
	m_{T3} &\in [200,7000] \ \text{GeV} \,, \\
	M_{1/2} &\in [200,7000] \ \text{GeV}
\end{align*}
and all other parameters fixed with values as shown in table~\ref{tab:univ-scan}. 
\begin{table}[h]
	\centering
	\begin{tabular}{lccccccccc}
		\toprule
		\textsc{Parameter} & $m_{F}$ & $m_{T1}$ & $m_{T2}$ & $m_{T3}$ & $M_{1/2}$ & $A_{\text{tri}}$ & $m_{H_{1,2}}$ & $\tan \beta$ & $\sgn \mu$ \\
		\midrule
		\textsc{Value}     & 6000  & 7000  & 300   & free  & free        & -6000            & 6000          & 30           & -1         \\
		\bottomrule
	\end{tabular}
	\caption{Input parameters at the GUT scale in GeV (apart for $\tan \beta$ and $\sgn \mu$) for universal gaugino masses $M_{1/2}$.}
	\label{tab:univ-scan}
\end{table}
An overview over the scanned $m_{T3}$-$M_{1/2}$ plane is shown in figure~\ref{fig:m3_vs_M3_vs_gmuon-univ}, where the colour coding indicates the value of $\Damu$.
The first thing to notice is that only a narrow stripe in the parameter space leads to radiative electroweak symmetry breaking (REWSB). Following the stripe to larger $m_{T3}$ and smaller $M_{1/2}$ gives larger $\Damu$, before the stripe eventually ends in a narrow peak around $(m_{T3},M_3) = (5.3,1.3)$ TeV. However, even in the peak $\Damu$ only reaches values up to $1.8 \times 10^{-10}$, which is about 10-20 times lower than observed. Before giving an explanation for why $\Damu$ is so small even with the assumptions made before, let us investigate the relic density and $\mu$ behaviour shown in figure~\ref{fig:gmuon_vs_Omega_univ}. 
Regarding the relic density shown in the left panel of figure~\ref{fig:gmuon_vs_Omega_univ}, it turns out that DM is mostly higgsino-like, thus yielding relic densities in the right range or maximally two orders of magnitude smaller than the observed upper limit. With increasing $\Damu$, the relic density slightly converges to some central value between its minimum and maximum reach. While the relic density thus is not a problem with this setup, the predicted DM DD cross sections turn out to be fully excluded (see colour-coding). This can be readily understood since dark matter in this case is dominantly higgsino-like and therefore has a significant coupling to the Higgs boson, leading to a large DM DD cross-section.

The right panel of figure~\ref{fig:gmuon_vs_Omega_univ} shows $\Damu$ as a function of $\mu$ and it turns out that smaller values of $\mu$ yield larger values of $\Damu$, as was expected (see section~\ref{sec:g-2} and the beginning of this section~\ref{sec:results}). It is also worth noticing that decreasing $\mu$ results in a decreased relic density due to the DM becoming more and more higgsino-like, as indicated by the colour-coded $\Omega h^2$.

In summary, the case of universal $M_{1/2}$ at the GUT scale with negative $\mu$ does not yield any values of $\Damu$ in or close to the 1$\sigma$ reference bound. This is expected and can be reasoned by the following argument. With negative $\mu$, only equations~\ref{eq:loops:B} and \ref{eq:loops:D} give positive contributions to $\Damu$, while the major differences between (\ref{eq:loops:B}) and (\ref{eq:loops:D}) are simply the exchange of $M_1$ and $M_2$ as well as $m_{\tilde{\mu}_R}$ and $m_{\tilde{\mu}_L}$. Since the loop functions only run from 0 to 1, they are irrelevant for our argument and we can conservatively assume for the moment that they both equal 1 and consider just the remaining prefactors. With $M_{1}$ and $M_2$ unified at the GUT scale, their low scale values will not be much different either and allow us to focus solely on one of the two equations, e.g. (\ref{eq:loops:B}). To get suitable $\Damu$, $M_1$ as well as $\mu$ need to be small ($\mathcal{O}(200)$ GeV). However, having $M_1$ that light will result in a similar light $M_3$ leading to light gluinos with masses $m_{\tilde{g}} \lesssim 1$ TeV~\cite{Baer:2004xx}. These are already excluded by LHC searches \cite{TheATLAScollaboration:2015cyl,Sirunyan:2017kqq} and hence lead to a contradiction. Additionally, too light $M_{1/2}$ will prevent REWSB from happening, as can be seen in figure~\ref{fig:m3_vs_M3_vs_gmuon-univ}.

Overall, in case of unified gaugino masses $M_{1/2}$, we did not find a region in parameter space able to explain $\Damu$ in harmony with the other experimental constraints considered. However, a possible solution arises when the gaugino masses are split into $M_{1,2}$ and $M_3$, allowing for heavy gluinos and light enough $M_{1,2}$ to yield the correct $\Damu$. This setup is studied in detail in the following section~\ref{sec:results-nonuniversal}.

\subsection{Partially Non-Universal Gaugino Masses}
\label{sec:results-nonuniversal}
\begin{figure}
	\centering
	\includegraphics[width=0.49\linewidth]{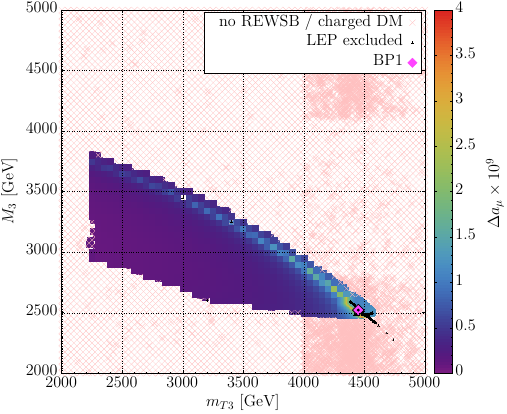}
	\hfill
	\includegraphics[width=0.49\linewidth]{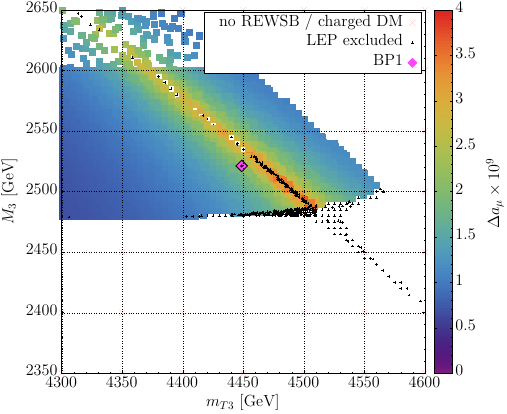}
	\caption{$m_{T3}$-$M_3$ plane with colour-coded $\Damu$ with non-universal gaugino masses. The panel at the right is an excerpt of the full scan shown in the left panel.}
	\label{fig:m3_vs_M3_vs_gmuon}
\end{figure}

Splitting $M_{1/2}$ into $M_{1,2}$ and $M_3$ allows us to keep $M_3$ heavy, while fixing $M_{1,2}$ to some value light enough to strengthen rather than weaken $\Damu$. We performed a scan taking this into account with
\begin{align*}
	m_{T3} &\in [500,7000] \ \text{GeV} \,, \\
	M_3 &\in [500,7000] \ \text{GeV}
\end{align*}
and all other parameters fixed with values as shown in table~\ref{tab:nonuniv-scan}.
%
%
\begin{table}[h]
	\centering
	\begin{tabular}{lcccccccccc}
		\toprule
		\textsc{Parameter} & $m_{F}$ & $m_{T1}$ & $m_{T2}$ & $m_{T3}$ & $M_{1,2}$ & $M_3$ & $A_{\text{tri}}$ & $m_{H_{1,2}}$ & $\tan \beta$ & $\sgn \mu$ \\
		\midrule
		\textsc{Value}     & 6000  & 7000  & 300   & free  & 250       & free  & -5000            & 6000          & 30           & -1         \\
		\bottomrule
	\end{tabular}
	\caption{Input parameters at the GUT scale in GeV for non-universal gaugino masses $M_{1,2}$ and $M_3$.}
	\label{tab:nonuniv-scan}
\end{table}
Analogue to figure~\ref{fig:m3_vs_M3_vs_gmuon-univ}, we show the scanned over $m_{T3}$-$M_3$ plane in figure~\ref{fig:m3_vs_M3_vs_gmuon}.
Similar to the case of universal gaugino masses, a narrow, slightly elliptic stripe of solutions with larger $\Damu$ can be seen for $M_3 \lesssim 3.8$ TeV and $m_{T3} \lesssim 4.5$ TeV. Additionally, a wide band around this stripe holds points where REWSB is happening, but $\Damu$ is close to zero. A second set of points with vanishingly small $\Damu$ is found for $M_3 \gtrsim 3$ TeV and $m_{T3} \gtrsim 6.5$ TeV (not shown here).
When zooming in on the interesting part of the scan with larger values of $\Damu$ (see right panel of figure~\ref{fig:m3_vs_M3_vs_gmuon}), we notice that the stripe extends into the nonphysical region without REWSB, although the points here are excluded by LEP limits due to too light charginos or smuons. Just before hitting the unphysical region, $\Damu$ peaks at values around $4 \times 10^{-9}$ before eventually vanishing abruptly in the non-REWSB region.
Comparing these first results to the case with universal gaugino masses, the large increase in $\Damu$ immediately becomes visible, therefore validating our assumptions made earlier.

\begin{figure}
	\centering
	\includegraphics[width=0.49\linewidth]{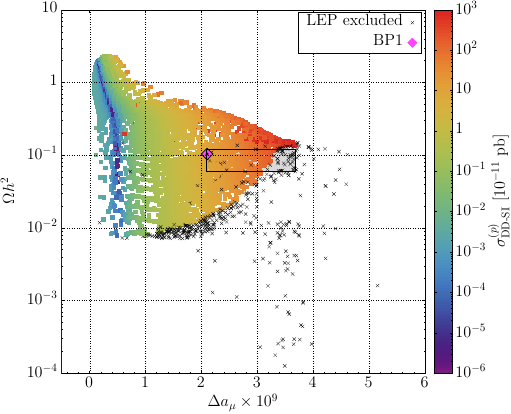}
	\hfill
	\includegraphics[width=0.49\linewidth]{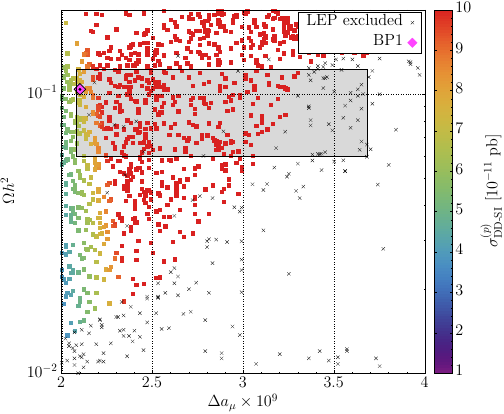}
	\caption{Relic density vs. $\Damu$ with colour-coded $\sigma^{(p)}_{\text{DD-SI}}$ with non-universal gaugino masses. The grey shaded rectangle shows the (extended) 1$\sigma$ bound for $\Damu$ ($\Omega h^2$). The panel at the right is an excerpt of the full scan shown in the left panel.}
	\label{fig:gmuon_vs_Omega}
\end{figure}

In figure~\ref{fig:gmuon_vs_Omega}, we show the relic density-$\Damu$ plane with colour-coded DM direct detection cross sections, analogue to figure~\ref{fig:gmuon_vs_Omega_univ}, left.
This time, however, dark matter is mainly bino-like and $\sigma_{\text{DD-SI}}$ is smaller than in figure~\ref{fig:gmuon_vs_Omega_univ} and increases faster with increasing $\Damu$. In the right panel of figure~\ref{fig:gmuon_vs_Omega}, a zoomed excerpt without logarithmic scaling\footnote{To allow for an easier comparison in the relevant range of $\sigma_{\text{DD-SI}}$, i.e. approximately between $1 \times 10^{-11}$ pb and $7.64 \times 10^{-11}$ pb, values of $\sigma_{\text{DD-SI}} > 10 \times 10^{-11}$ pb are also coloured red.} of $\sigma_{\text{DD-SI}}$ shows that most of the 1$\sigma$ reference bounds for $\Damu$ and $\Omega h^2$ is excluded by DM direct detection, only leaving a small range of solutions for the lower edge of the $\Damu$ 1$\sigma$ bound. Nevertheless, in comparison to universal gaugino masses, there are solutions for non-universal gaugino masses that satisfy all experimental limits.

\begin{figure}
	\centering
	\includegraphics[width=0.49\linewidth]{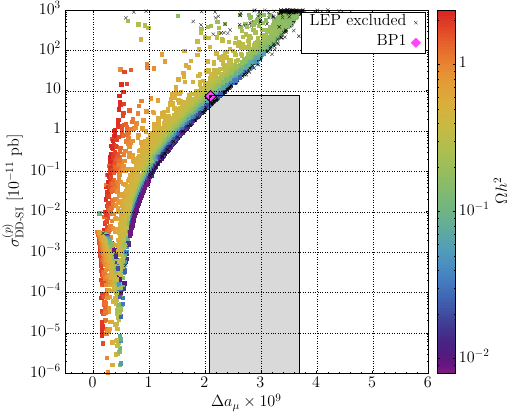}
	\hfill
	\includegraphics[width=0.49\linewidth]{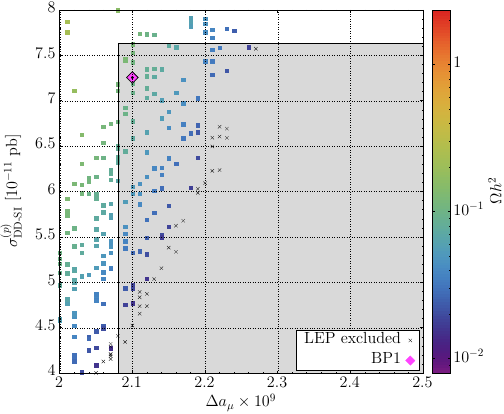}
	\caption{$\sigma_{\text{DD-SI}}$ vs. $\Damu$ with colour-coded relic density $\Omega h^2$ with non-universal gaugino masses. The grey shaded rectangle shows the 1$\sigma$ bound for $\Damu$ and the upper limit for $\sigma_{\text{DD-SI}}$. The panel at the right is an excerpt of the full scan shown in the left panel.}
	\label{fig:gmuon_vs_DD}
\end{figure}

Similar to figure~\ref{fig:gmuon_vs_Omega}, figure~\ref{fig:gmuon_vs_DD} holds the same data but with $\Omega h^2$ and $\sigma_{\text{DD-SI}}$ switched. Presenting the data this way allows for a better understanding of the excluded and allowed parameter space with respect to $\sigma_{\text{DD-SI}}$. As can be seen in figure~\ref{fig:gmuon_vs_DD}, right, only a small fraction of points falls within the 1$\sigma$ reference bounds of $\Damu$ and $\sigma_{\text{DD-SI}}$ (grey rectangle), although the majority of these points provides a very good relic density.

\begin{figure}
	\centering
	\includegraphics[width=0.49\linewidth]{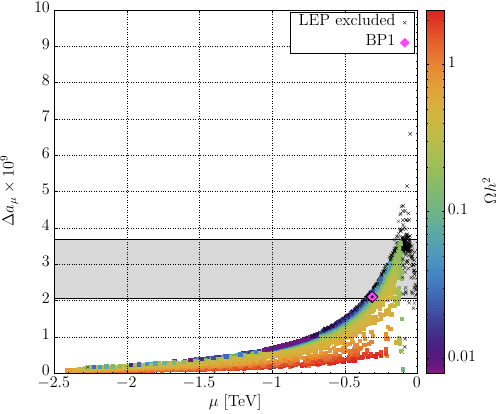}
	\hfill
	\includegraphics[width=0.49\linewidth]{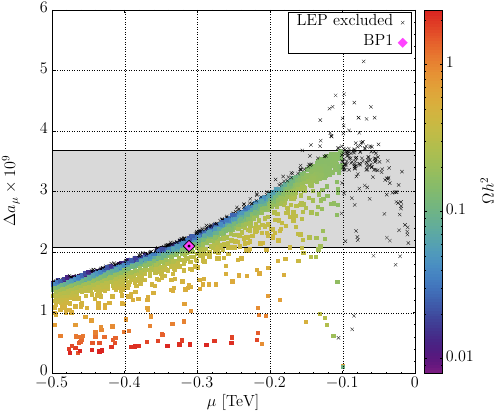}
	\caption{$\Damu$ vs. $\mu$ with colour-coded relic density $\Omega h^2$ with non-universal gaugino masses. The grey shaded rectangle shows the 1$\sigma$ bound for $\Damu$. The panel at the right is an excerpt of the full scan shown in the left panel.}
	\label{fig:mu_vs_gmuon}
\end{figure}

In figure~\ref{fig:mu_vs_gmuon}, the $\mu$ dependence of $\Damu$ is shown and it turns out that $\mu$ needs to be between $-300$ GeV and $-100$ GeV in order to yield the desired $\Damu$. When $\mu$ goes closer to 0, the higgsino components of the LSP start to dominate while simultaneously, the mass of the lightest chargino falls below approximately 100 GeV. Such light charginos are excluded by LEP \cite{Pasztor:2005es}, thus limiting our parameter space to values of $\mu$ smaller than $-100$ GeV.

\begin{figure}
	\centering
	\includegraphics[width=0.49\linewidth]{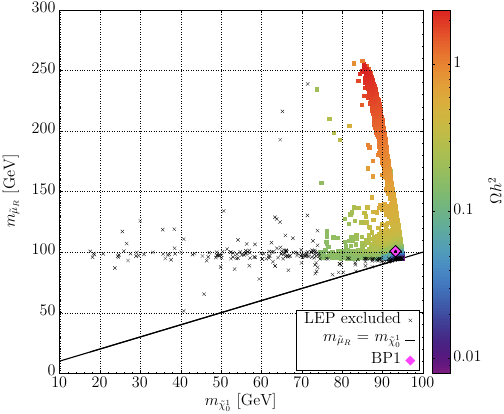}
	\hfill
	\includegraphics[width=0.49\linewidth]{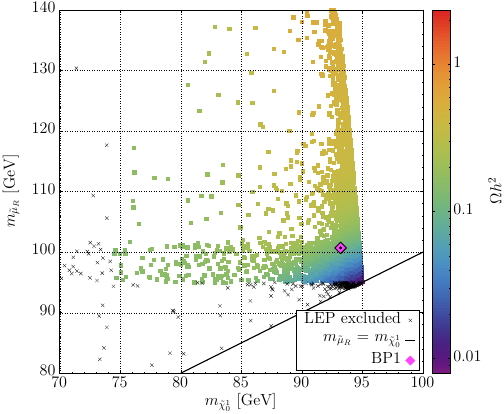}
	\caption{$m_{\tilde{\mu}_R}$ vs. $m_{\tilde{\chi}^0_1}$ with colour-coded relic density $\Omega h^2$ with non-universal gaugino masses. The panel at the right is an excerpt of the full scan shown in the left panel.}
	\label{fig:msmuR_vs_mneut1}
\end{figure}

In figure~\ref{fig:msmuR_vs_mneut1}, we show the $m_{\tilde{\mu}_R}$-$m_{\tilde{\chi}^0_1}$ plane with colour-coded relic density. As can be seen in the right panel, the pink benchmark point sits well above the line where the right-handed smuon and LSP are mass-degenerate. For this benchmark point, the LSP is predominantly bino-like, but with a non-zero higgsino component. This allows for a significant amount of $\tilde{\chi}^0_1$-$\tilde{\chi}^0_1$ annihilation in addition to the dominant $\tilde{\mu}_R$-$\tilde{\chi}^0_1$ co-annihilation cross-section leading to the correct relic density.

\begin{figure}
	\centering
	\includegraphics[width=0.49\linewidth]{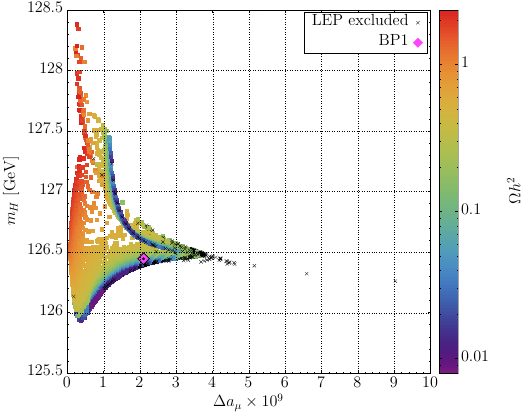}
	\hfill
	\includegraphics[width=0.49\linewidth]{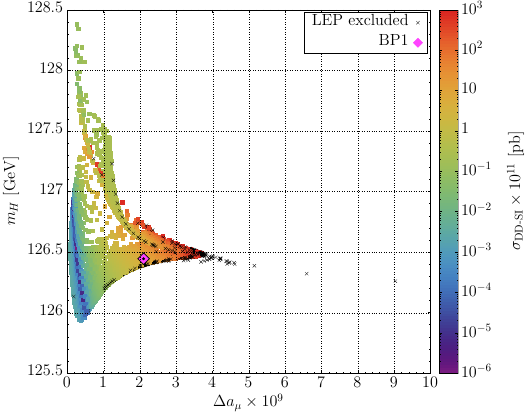}
	\caption{$m_h$ vs. $\Damu$ with colour-coded $\Omega h^2$ (left) and $\sigma_{\text{DD-SI}}$ (right) with non-universal gaugino masses.}
	\label{fig:mH_vs_gmuon}
\end{figure}

In figure~\ref{fig:mH_vs_gmuon}, we show the Higgs mass $m_h$ as a function of $\Damu$ with colour-coded $\Omega h^2$ (left) and $\sigma_{\text{DD-SI}}$ (right). For small values of $\Damu$, a broad range of Higgs masses is accessible with REWSB. This range shrinks drastically with increasing $\Damu$ and eventually peaks at $m_h = 126.5$ GeV for $\Damu \approx 4 \times 10^{-9}$. The relic density generally decreases with increasing $\Damu$, while the DM DD cross sections increase, as discussed before.

\begin{figure}
	\centering
	\includegraphics[width=0.60\linewidth]{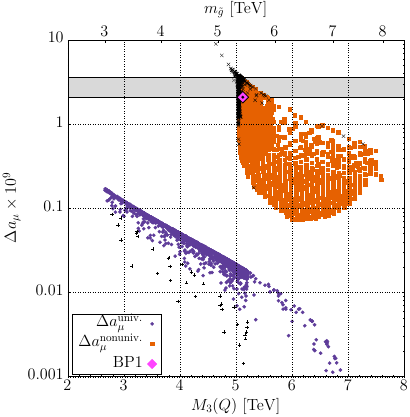}
	\caption{Influence of having universal (non-universal) gaugino masses $M_{1/2}$ $(M_{1,2}, M_3)$ on $\Damu$. The purple (red) points represent the universal (non-universal) case. The grey shaded rectangle shows the 1$\sigma$ bound for $\Damu$. Note that, to allow for an easier comparison, the non-universal points were gathered with $A_{\text{tri}} = -6$ TeV instead of $A_{\text{tri}} = -5$ TeV as shown for the figures~\ref{fig:m3_vs_M3_vs_gmuon} - \ref{fig:msmuR_vs_mneut1}.}
	\label{fig:comparison}
\end{figure}

Lastly, in figure~\ref{fig:comparison} in the right panel we show a comparison between $\Damu$ as a function of $M_3(Q)$ (lower horizontal axis) and $m_{\tilde{g}}$ (top horizontal axis) for both universal (purple diamonds) and non-universal (orange squares) gaugino masses. It is clearly visible that universal gaugino masses canot lead to viable $\Damu$ and --- even if there were a way to increase $\Damu$ further --- the gluinos would become quite light, potentially violating existing collider constraints. In case of non-universal gaugino masses, the $\Damu$ spectrum with respect to $M_3$ is slightly squeezed, but approximately one order of magnitude larger. This leads to a large spectrum of points with $\Damu$ in the correct range while simultaneously keeping the gluinos fairly heavy.
Overall, having non-universal gaugino masses allows for a variety of points with viable $\Damu$, which then can be tested further against other experimental constraints, as was shown above. Based on these findings, we provide three qualitatively different benchmark points, summarised in table~\ref{tab:benchmark_tab} below. BP2 differs from BP1 mainly in having $\tan \beta = 28$ and $A_{\text{tri}} = 0$, whereas BP3 has a non-vanishing negative $A_{\text{tri}}$ and split $m_F$ and $m_{T1}$.
\begin{table}[htbp]
	\centering
    \resizebox{0.6\textwidth}{!}{%
	\begin{tabular}{cccccc}
\toprule
 & \textsc{Benchmark}: & BP1 & BP2 & BP3 & \\ 
\midrule
\multirow{11}{*}{\rotatebox{90}{\textsc{Input at GUT scale}}} & $\tan \beta$ & 30 & 28 & 28 & \\
 & sgn$(\mu)$       & -       & -       & -       & \\\cline{2-6}
 & $m_F$            & 6000.0  & 6000.0  & 6200.0  & \multirow{10}{*}{\rotatebox{-90}{[GeV]}} \\
 & $m_{T1}$         & 7000.0  & 6000.0  & 5700.0  & \\
 & $m_{T2}$         & 300.0   & 300.0   & 290.0   & \\
 & $m_{T3}$         & 4448.6  & 5572.0  & 5518.0  & \\
 & $M_{1,2}$        & 250.0   & 250.0   & 250.0   & \\
 & $M_3$            & 2521.2  & 2446.0  & 2790.0  & \\
 & $M_{h_1}$        & 6000.0  & 6000.0  & 6200.0  & \\
 & $M_{h_2}$        & 6000.0  & 6000.0  & 6200.0  & \\
 & $A_{\text{tri}}$ & -5000.0 & 0       & -500.0  & \\
\hline
\multirow{27}{*}{\rotatebox{90}{\textsc{Masses}}} & $m_{h}$ & 126.4 & 124.3 & 124.7 &\multirow{29}{*}{\rotatebox{-90}{[GeV]}} \\
 & $m_{\tilde{g}}$          & 5457.7  & 5280.9  & 5963.4 & \\
 & $m_{\tilde{q}^1_L}$      & 8248.5  & 7312.5  & 7433.2 & \\
 & $m_{\tilde{u}_R}$        & 8250.1  & 7316.9  & 7439.2 & \\
 & $m_{\tilde{q}^2_L}$      & 4350.1  & 4173.2  & 4764.6 & \\
 & $m_{\tilde{c}_R}$        & 4377.1  & 4198.9  & 4788.7 & \\
 & $m_{\tilde{b}_1}$        & 4866.7  & 5884.2  & 6162.0 & \\
 & $m_{\tilde{t}_1}$        & 3944.4  & 5068.5  & 5340.8 & \\
 & $m_{\tilde{t}_2}$        & 4875.0  & 5887.4  & 6165.7 & \\
 & $m_{\tilde{d}_R}$        & 7423.9  & 7320.6  & 7832.1 & \\
 & $m_{\tilde{s}_R}$        & 7423.8  & 7320.5  & 7831.9 & \\
 & $m_{\tilde{b}_2}$        & 6934.5  & 6947.4  & 7453.3 & \\
 & $m_{\tilde{e}_L}$        & 5987.1  & 5988.4  & 6188.8 & \\
 & $m_{\tilde{e}_R}$        & 7001.2  & 5999.3  & 5699.4 & \\
 & $m_{\tilde{\mu}_L}$      & 5986.5  & 5988.0  & 6188.3 & \\
 & $m_{\tilde{\mu}_R}$      & 100.7   & 95.6    & 95.4   & \\
 & $m_{\tilde{\tau}_1}$     & 3731.8  & 5175.0  & 5057.0 & \\
 & $m_{\tilde{\tau}_2}$     & 5737.5  & 5789.7  & 5989.0 & \\
 & $m_{\tilde{\chi}^0_1}$   & 93.2    & 91.1    & 89.2   & \\
 & $m_{\tilde{\chi}^0_2}$   & 169.4   & 163.6   & 158.7  & \\
 & $m_{\tilde{\chi}^0_3}$   & -341.9  & -336.2  & -337.8 & \\
 & $m_{\tilde{\chi}^0_4}$   & 353.9   & 347.8   & 348.6  & \\
 & $m_{\tilde{\chi}^\pm_1}$ & 169.6   & 163.7   & 158.9  & \\
 & $m_{\tilde{\chi}^\pm_2}$ & 356.8   & 350.7   & 351.5  & \\
 & $m_{\tilde{\nu}^e_L}$    & 5986.1  & 5987.5  & 6187.8 & \\
 & $m_{\tilde{\nu}^\mu_L}$  & 5985.6  & 5987.0  & 6187.3 & \\
 & $m_{\tilde{\nu}^\tau_L}$ & 5736.8  & 5788.7  & 5988.1 & \\\cline{1-5}
 & $Q$                      & 4287.9  & 5353.0  & 5609.8 & \\
 & $\mu$                    & -311.5  & -302.1  & -299.5 & \\
\hline
\multirow{5}{*}{\rotatebox{90}{\textsc{Constraints}}} & Br$(b \to s \gamma)$      & $3.40 \times 10^{-4}$  & $3.35 \times 10^{-4}$  & $3.34 \times 10^{-4}$ & \\
 & Br$(B_s \to \mu^+ \mu^-)$ & $3.03 \times 10^{-9}$  & $3.04 \times 10^{-9}$  & $3.04 \times 10^{-9}$  &  \\
 & $\sigma^{\text{DD SI}}$   & $7.23 \times 10^{-11}$ & $7.59 \times 10^{-11}$ & $6.89 \times 10^{-11}$ & [pb]  \\
 & $\Omega h^2$              & $1.04 \times 10^{-1}$  & $4.65 \times 10^{-2}$  & $7.55 \times 10^{-2}$  &  \\
 & $\Delta a_\mu$            & $2.10 \times 10^{-9}$  & $2.09 \times 10^{-9}$  & $2.09 \times 10^{-9}$  &  \\
\bottomrule
    \end{tabular}}
    \caption{Input and output parameters for the benchmark points with partial gaugino non-universality $M_1=M_2\ll M_3$.
    These points have good $\Delta a_\mu$ as well as $\Omega h^2$ but the wino dominated charginos ${\tilde{\chi}^\pm_1}$ and neutralinos 
    ${\tilde{\chi}^0_2}$ are too light to have 
    avoided 8 TeV LHC searches as discussed in the text. $\tilde{q}^{i}$ labels the $i$-th generation of squarks.}
    \label{tab:benchmark_tab}
\end{table}

The benchmark points in this region are characterised by:
a) bino dominated $\tilde{\chi}^0_1$ LSP being the Dark Matter particle with a mass below about 100 GeV; b) a next-to-lightest
right-handed smuon $\tilde\mu_R$ with mass several GeV heavier; c) wino dominated $\tilde{\chi}^0_2$ and $\tilde{\chi}^\pm_1$ having a mass gap between them and $\tilde{\chi}^0_1$ of less than the $Z$ or $W$ boson masses respectively; 
d) non-negligible $\tilde{\mu}_R -\tilde{\mu}_L$ mixing (enhanced by not-so-small values of $\tan\beta$) and respectively non-negligible $\tilde\chi^\pm_1 \to \tilde{\mu}^\pm_R \, \nu_\mu$ decay branching fractions;
e) higgsino dominated $\tilde{\chi}^0_3$ and $\tilde{\chi}^\pm_2$ with masses below 400 GeV;
f) all other SUSY partners having multi-TeV masses. 

Such a specific spectrum of light electroweak gauginos and right-handed smuons predicts a rather characteristic signal at the LHC. 
The signal comes dominantly from $\tilde{\chi}^\pm_1\tilde{\chi}^0_2$ and $\tilde{\chi}^+_1\tilde{\chi}^-_1$-pair production followed by the dominant $\tilde{\chi}^0_2$ decay into a smuon which  --- in its turn --- decays into a muon and DM. On the other hand, due to the non-negligible $\tilde{\mu}_R$-$\tilde{\mu}_L$ mixing mentioned above, the branching ratio for $\tilde\chi^\pm \to \tilde{\mu}^\pm_R \, \nu_\mu$  becomes comparable to the 3-body decay $\tilde\chi^\pm_1 \to f\bar{f'} \, \tilde\chi^0_1$ via a virtual $W$ boson. This $Br(\tilde{\chi}^\pm_1\to {\tilde{\mu}}^\pm \, \nu_\mu)$ can be   substantial ($\simeq 30$-$50\%$) 
 because of the significant higgsino component. The signal strength $m_{\tilde{\mu}^\pm}$ strongly depends on the $\tilde{\mu}_R$-$\tilde{\chi}^0_1$
 mass gap and  can be quite hidden if this mass gap is small (below a few GeV) since in this case the smuon decay products will be soft.
 The $\tilde{\chi}^0_2$ decay is characterised by the dominant $\tilde{\chi}^0_2 \to \tilde{\mu}_R \, \nu_\mu$ decay with not-so-soft leptons
 (energy of which is independent of $\tilde{\mu}_R$-$\tilde{\chi}^0_1$ mass gap)
 providing a very important contribution to the leptonic signature.
 Thus, the only signature from the scenario under study is very specific and characterised by muon-dominated di- and tri- lepton signatures at the LHC.

We have performed a \texttt{CheckMATE 2.0.11} \cite{Dercks:2016npn} analysis on these three benchmark points,
including all implemented 8 and 13 TeV ATLAS and CMS analyses on chargino and neutralino searches with 
a light smuon
and have verified that the LHC in fact is highly sensitive to this part of the parameter space.
In particular, we used \texttt{MadGraph 5.2.3.3} \cite{Alwall:2014hca} linked to \texttt{CheckMATE} to generate 50000 events for SUSY final states consisting of $\tilde{\mu}^\pm_R$, $\tilde{\chi}^0_1$, $\tilde{\chi}^0_2$ as well as $\tilde{\chi}^\pm_1$. Next, \texttt{PYTHIA 8.2.30} \cite{Sjostrand:2007gs} was used to shower and hadronise the events and eventually \texttt{CheckMATE} together with \texttt{Delphes 3.3.3} \cite{deFavereau:2013fsa} was used to perform the event and detector analysis. While setting the same cuts as were used for the experimental analyses, the \texttt{CheckMATE} framework therefore allows us to examine whether given points in the parameter space are allowed or ruled out by current experimental searches. 
For all three benchmarks, the ATLAS search \texttt{ATLAS\_1402\_7029} \cite{Aad:2014nua} aimed at three leptons plus missing $E_T$ was most sensitive. The $r_\text{max}$ value defined by \cite{Dercks:2016npn}
\begin{equation}
  r_\text{max} = \frac{S - 1.64 \cdot \Delta S}{S95}\,, \notag
  \label{eq:r-value}
\end{equation}
where $S$ is the number of predicted signal events with its uncertainty $\Delta S$ and $S95$ is the experimental 95 \% upper limit on the number of signal events, is shown below in table~\ref{tab:rvalBP} for all three benchmarks. Values of $r_\text{max} \geq 1$ indicate the signal is excluded, whereas $r_\text{max} < 1$ indicates that the signal is not excluded or probed yet.
\begin{table}[h]
	\centering
    \resizebox{0.99\textwidth}{!}{%
	\begin{tabular}{lcccc}
		\toprule
		\multirow{2}{*}{\textsc{Quantity}} & \multirow{2}{*}{\textsc{Unit}} & \multicolumn{3}{c}{\textsc{Benchmark}} \\
		                                                                              &         & BP1                        & BP2                             & BP3                             \\
		\midrule
		$r_\text{max}$                                                                &         & 7.38                       & 9.16                             & 9.30                            \\
		$\sqrt{s}$                                                                    & TeV     & 8                          & 8                             & 8                             \\
		\textsc{Analysis}                                                             &         & \texttt{ATLAS\_1402\_7029} & \texttt{ATLAS\_1402\_7029} & \texttt{ATLAS\_1402\_7029} \\
		\textsc{Signal Region}                                                        &         & \texttt{SR0taua06}         & \texttt{SR0taua02}                  & \texttt{SR0taua02}                    \\
		\textsc{Ref.}                                                                 &         & \cite{Aad:2014nua}         & \cite{Aad:2014nua}            & \cite{Aad:2014nua}            \\
		$\sigma_{\text{LO}}$                                                          & pb      & 1.65                       & 1.85                            & 2.14                            \\
		\midrule
		BR$(\tilde{\chi}^0_2 \to \tilde{\mu}^\pm_R \, \mu^\mp)$                       & \%      & 99.4                       & 99.4                            & 99.7                           \\
		BR$(\tilde{\chi}^0_2 \to \bar{q} \, q \, \tilde{\chi}^0_1)$                   & \%      & 0.4                        & 0.4                            & 0.2                            \\
		BR$(\tilde{\chi}^0_2 \to \ell^\pm \, \ell^\mp \, \tilde{\chi}^0_1)$           & \%      & 0.1                        & 0.1                            & $< 0.1$                            \\
		BR$(\tilde{\chi}^0_2 \to \bar{\nu}_\ell \, \nu_\ell \, \tilde{\chi}^0_1)$     & \%      & $< 0.1$                    & $< 0.1$                        & $< 0.1$                            \\
		\midrule
		BR$(\tilde{\chi}^\pm_1 \to \bar{d}^{1,2} \, u^{1,2} \, \tilde{\chi}^0_1)$     & \%      & 45.4                       & 40.2                           & 47.9                           \\		
		BR$(\tilde{\chi}^\pm_1 \to \tilde{\mu}^\pm_R \, \nu_\mu)$                     & \%      & 31.9                       & 39.8                           & 34.7                       \\		
		BR$(\tilde{\chi}^\pm_1 \to \ell^\pm \, \nu_\ell \, \tilde{\chi}^0_1)$         & \%      & 22.7                       & 20.0                           & 17.4                           \\		
		\midrule		
		$\Delta m(\tilde{\chi}^\pm_1,\tilde{\mu}_R)$                                  & GeV     & 68.9                       & 67.9                           & 63.5                           \\
		$\Delta m(\tilde{\chi}^0_2,\tilde{\mu}_R)$                                    & GeV     & 68.7                       & 67.7                           & 63.3                           \\
		$\Delta m(\tilde{\mu}_R,\tilde{\chi}^0_1)$                                    & GeV     & 7.5                        & 6.6                            & 6.2                           \\
		\bottomrule
	\end{tabular}}
	\caption{\texttt{CheckMATE} analysis results for the benchmarks of table~\ref{tab:benchmark_tab}
	with partial gaugino non-universality $M_1=M_2\ll M_3$.}
	\label{tab:rvalBP}
\end{table}

It turns out that all benchmarks are strongly excluded, which is mainly due to the light $\tilde{\chi}^\pm_1$ and $\tilde{\chi}^0_2$ and their subsequent decays to the right-handed smuon. 

A summary of the most powerfully excluding  LHC  searches for BP1 -- BP3 is given
in table~\ref{tab:rvalBP}, where we present the $r_\text{max}$ value from \texttt{CheckMATE}
together with properties of the principal decay channels for $\tilde\chi^\pm_1$ and 
$\tilde\chi^0_2$. 
The most sensitive search is actually done by ATLAS for the 8 TeV data \texttt{ATLAS\_1402\_7029}~\cite{Aad:2014nua}
and the most sensitive  signature is the tri-lepton one,
containing always  one soft muon from the $\tilde{\mu}_R \to \tilde{\chi}_1^0 \, \mu$ decay.
Even though this muon is soft, the well designed asymmetric $p_T$ cuts for the leptons
in Ref.~\cite{Aad:2014nua} allow for being sensitive to a second or third lepton with $p_T$ 
as low as 10 GeV. To the best of our knowledge, analogue 13 TeV searches are not sensitive
to such low $p_T$ leptons.

\begin{table}[htbp]
	\centering
    \resizebox{0.6\textwidth}{!}{%
	\begin{tabular}{cccccc}
\toprule
 & \textsc{Benchmark}: & BP4 & BP5 & BP6 & \\ 
\midrule
\multirow{12}{*}{\rotatebox{90}{\textsc{Input at GUT scale}}} & $\tan \beta$ & 30 & 28 & 30 & \\
 & sgn$(\mu)$       & -       & -       & -       & \\\cline{2-6}
 & $m_F$            & 5000.0  & 6200.0 & 5000.0  & \multirow{10}{*}{\rotatebox{-90}{[GeV]}} \\
 & $m_{T1}$         & 5000.0  & 5700.0 & 5000.0  & \\
 & $m_{T2}$         & 200.0   & 280.0  & 200.0   & \\
 & $m_{T3}$         & 2995.0  & 5430.0 & 3005.0  & \\
 & $M_1$            & 250.0   & 250.0  & 250.0   & \\
 & $M_2$            & 400.0   & 550.0  & 500.0   & \\
 & $M_3$            & 2600.0  & 2945.0 & 2595.0  & \\
 & $M_{h_1}$        & 5000.0  & 6200.0 & 5000.0  & \\
 & $M_{h_2}$        & 5000.0  & 6200.0 & 5000.0  & \\
 & $A_{\text{tri}}$ & -4000.0 & -500.0 & -4000.0 & \\
\hline
\multirow{27}{*}{\rotatebox{90}{\textsc{Masses}}} & $m_{h}$ & 126.3 & 124.7 & 126.2 &\multirow{29}{*}{\rotatebox{-90}{[GeV]}} \\
 & $m_{\tilde{g}}$          & 5531.7 & 6235.3 & 5516.5 & \\
 & $m_{\tilde{q}^1_L}$      & 6743.0 & 7589.2 & 6735.7 & \\
 & $m_{\tilde{u}_R}$        & 6743.7 & 7589.9 & 6734.1 & \\
 & $m_{\tilde{q}^2_L}$      & 4516.4 & 5003.3 & 4505.7 & \\
 & $m_{\tilde{c}_R}$        & 4529.2 & 5018.0 & 4514.9 & \\
 & $m_{\tilde{b}_1}$        & 4312.4 & 6262.8 & 4306.4 & \\
 & $m_{\tilde{t}_1}$        & 3601.6 & 5443.3 & 3588.2 & \\
 & $m_{\tilde{t}_2}$        & 4324.0 & 6266.7 & 4318.0 & \\
 & $m_{\tilde{d}_R}$        & 6748.0 & 7975.4 & 6738.4 & \\
 & $m_{\tilde{s}_R}$        & 6747.9 & 7975.3 & 6738.3 & \\
 & $m_{\tilde{b}_2}$        & 6348.2 & 7597.3 & 6337.5 & \\
 & $m_{\tilde{e}_L}$        & 4994.9 & 6196.1 & 4998.5 & \\
 & $m_{\tilde{e}_R}$        & 5002.1 & 5699.9 & 5002.1 & \\
 & $m_{\tilde{\mu}_L}$      & 4994.4 & 6195.6 & 4998.0 & \\
 & $m_{\tilde{\mu}_R}$      & 98.9   & 96.8   & 99.4   & \\
 & $m_{\tilde{\tau}_1}$     & 2282.9 & 4968.1 & 2293.7 & \\
 & $m_{\tilde{\tau}_2}$     & 4802.1 & 5999.4 & 4805.3 & \\
 & $m_{\tilde{\chi}^0_1}$   & 91.7   & 89.0   & 92.0   & \\
 & $m_{\tilde{\chi}^0_2}$   & 266.9  & 303.3  & 302.2  & \\
 & $m_{\tilde{\chi}^0_3}$   & -335.1 & -327.8 & -335.9 & \\
 & $m_{\tilde{\chi}^0_4}$   & 376.8  & 458.9  & 430.4  & \\
 & $m_{\tilde{\chi}^\pm_1}$ & 267.4  & 303.7  & 302.8  & \\
 & $m_{\tilde{\chi}^\pm_2}$ & 378.2  & 459.0  & 430.7  & \\
 & $m_{\tilde{\nu}^e_L}$    & 4993.8 & 6195.1 & 4997.4 & \\
 & $m_{\tilde{\nu}^\mu_L}$  & 4993.4 & 6194.6 & 4997.0 & \\
 & $m_{\tilde{\nu}^\tau_L}$ & 4800.9 & 5998.4 & 4804.1 & \\\cline{1-5}
 & $Q$                      & 3866.1 & 5705.8 & 3856.5 & \\
 & $\mu$                    & -313.0 & -293.3 & -314.3 & \\
\hline
\multirow{5}{*}{\rotatebox{90}{\textsc{Constraints}}} & Br$(b \to s \gamma)$      & $3.43 \times 10^{-4}$  & $3.34 \times 10^{-4}$  & $3.43 \times 10^{-4}$ & \\
 & Br$(B_s \to \mu^+ \mu^-)$ & $3.01 \times 10^{-9}$  & $3.04 \times 10^{-9}$  & $3.01 \times 10^{-9}$  &  \\
 & $\sigma^{\text{DD SI}}$   & $6.72 \times 10^{-11}$ & $6.81 \times 10^{-11}$ & $6.52 \times 10^{-11}$ & [pb]  \\
 & $\Omega h^2$              & $9.67 \times 10^{-2}$  & $1.10 \times 10^{-1}$  & $1.03 \times 10^{-1}$  &  \\
 & $\Delta a_\mu$            & $2.17 \times 10^{-9}$  & $2.14 \times 10^{-9}$  & $2.16 \times 10^{-9}$  &  \\
\bottomrule
    \end{tabular}}
    \caption{Input and output parameters for the benchmark points with full gaugino non-universality $M_1<M_2\ll M_3$.
These points have good $\Delta a_\mu$ as well as $\Omega h^2$ with all other constraints being fulfilled. In particular
the higgsino dominated charginos ${\tilde{\chi}^\pm_1}$ and neutralinos 
    ${\tilde{\chi}^0_2}$ are heavy enough to have 
    avoided current LHC searches, but are a target for future searches, as discussed in the text.
$\tilde{q}^{i}$ labels the $i$-th generation of squarks.}
    \label{tab:benchmark_tab-fully-nonuniv}
\end{table}

\subsection{Fully Non-Universal Gaugino Masses}
\label{sec:results-fully-nonuniversal}

So far, in the previous subsections we have shown that our scenario for the muon $g-2$ requires a light right-handed smuon around 100 GeV together with a neutralino several GeV lighter leading to successful dark matter. We have seen that such a scenario is not consistent with universal gauginos at the GUT scale due to the gluino mass bound, which requires $M_{1,2}\ll M_3$. We have also seen that this scenario is not consistent with $M_1=M_2$ due to the subsequent prediction of wino dominated charginos and neutralinos with masses around 160--170 GeV, which are excluded by 8 TeV LHC searches that are most sensitive for the resulting soft muons arising from smuon decays. 

Here we shall show that, allowing fully non-universal gaugino masses with 
$M_1<M_2 \ll M_3$, gives charginos and neutralinos which are somewhat heavier, thereby satisfying current LHC search constraints.
With such full non-universality, we may then access regions of parameter space where $M_2$ exceeds the magnitude of the higgsino mass parameter 
(typically $\mu\sim -300$ GeV as required to achieve a successful muon $g-2$).
Then, the charginos and neutralinos become higgsino dominated with masses governed by $|\mu|\sim 300$ GeV.
The full scans of the parameter space are quite analogous to those in the previous subsection, with the only difference
being that $M_2$ is somewhat heavier than $M_1$. Therefore it suffices to present a few new benchmark
points to illustrate the effect of having $M_1<M_2 \ll M_3$.

In table~\ref{tab:benchmark_tab-fully-nonuniv}, we define three new benchmark points BP4, BP5 and BP6, corresponding to having $M_1<M_2 \ll M_3$. 
The benchmark points in this region are characterised by:
a) bino dominated $\tilde{\chi}^0_1$ LSP being the Dark Matter particle with a mass below about 100 GeV;
b) a next-to-lightest right-handed smuon $\tilde\mu_R$ with a mass several GeV heavier;
c) higgsino dominated $\tilde{\chi}^0_2$ and $\tilde{\chi}^\pm_1$ with masses governed by $|\mu|\sim 300$ GeV; 
d) wino dominated $\tilde{\chi}^0_3$ and $\tilde{\chi}^\pm_2$ with masses governed by $M_2$;
e) all other SUSY partners having multi-TeV masses. 

The main difference from the previous benchmarks is that the wino dominated charginos and neutralinos are now pushed 
up in mass due to the increase in $M_2$. However, the remaining higgsino dominated charginos and neutralinos whose 
mass is governed by $|\mu|$ cannot be pushed up beyond $\simeq 300 $~GeV, since we need $\mu\sim -300$ GeV to achieve a successful muon $g-2$.
These charginos and neutralinos therefore remain a target for LHC searches.
We have again performed a \texttt{CheckMATE 2.0.11} analysis on these three benchmark points,
including all implemented 8 and 13 TeV ATLAS and CMS analyses on chargino and neutralino searches with 
a light smuon and have verified that the LHC in fact is highly sensitive to this part of the parameter space.
Following the procedure described in detail in the previous subsection, 
we have obtained the results shown in table~\ref{tab:rvalBP-fully-nonuniv} for all three benchmarks.
Contrary to the previous results, now we see that all three benchmark points are consistent with current LHC searches,
however BP4 is on the verge of being excluded with a value of $r_\text{max} = 0.88$,
while BP5 and BP6 both have $r_\text{max} \approx 0.12$ and will require a substantial increase in luminosity
to exclude them. The search channels are di- and tri-lepton searches plus missing energy, as before,
but since the chargino and neutralino masses are larger, the cross-sections are now lower,
as can be seen in table~\ref{tab:rvalBP-fully-nonuniv}.

Another reason why the sensitivity of the LHC to BP4 -- BP6 is lower
in comparison to the BP1 -- BP3 case is because of the new decay 
channel $\tilde{\chi}^0_2 \to h \, \tilde{\chi}^0_1$
to which the current LHC searches have lower sensitivity.
One can see from table~\ref{tab:rvalBP-fully-nonuniv} that the
branching ratio to this channel is substantial (about 50 \%),
which eventually further lowers the LHC sensitivity.
One should also note that BP5 and BP6 represent the region of the parameter space 
to which the LHC is currently the least sensitive.
Nevertheless, with a future total integrated luminosity of about 3 ab$^{-1}$, the 
LHC will be able to probe even these corners of the parameter space
with di- and tri-lepton signatures from higgsino production.
Moreover, the increase of sensitivity of the DM direct detection experiments
by a factor of two, which is expected to take place in the next few years,
will independently probe the entire parameter space of the scenario under study.

\begin{table}[h]
	\centering
    \resizebox{0.99\textwidth}{!}{%
	\begin{tabular}{lcccc}
		\toprule
		\multirow{2}{*}{\textsc{Quantity}} & \multirow{2}{*}{\textsc{Unit}} & \multicolumn{3}{c}{\textsc{Benchmark}} \\
		                                   &                                & BP4                             & BP5                             & BP6                             \\
		\midrule
		$r_\text{max}$                     &                                & 0.88                            & 0.12                            & 0.13                            \\
		$\sqrt{s}$                         & TeV                            & 13                              & 13                              & 13                              \\
		\textsc{Analysis}                  &                                & \texttt{ATLAS\_CONF\_2016\_096} & \texttt{ATLAS\_CONF\_2016\_096} & \texttt{ATLAS\_CONF\_2016\_096} \\
		\textsc{Signal Region}             &                                & \texttt{3LI}                    & \texttt{2LADF}                  & \texttt{3LI}                    \\
		\textsc{Ref.}                      &                                & \cite{ATLAS:2016uwq}            & \cite{ATLAS:2016uwq}            & \cite{ATLAS:2016uwq}            \\
		$\sigma_{\text{LO}}$               & pb                             & 0.54                            & 0.24                            & 0.26                            \\
		\midrule
		BR$(\tilde{\chi}^0_2 \to h \, \tilde{\chi}^0_1)$          & \%      & 51.0                            & 55.5                            & 55.4                            \\
		BR$(\tilde{\chi}^0_2 \to Z \, \tilde{\chi}^0_1)$          & \%      & 30.5                            & 30.2                            & 30.1                            \\
		BR$(\tilde{\chi}^0_2 \to \tilde{\mu}^\pm_R \, \mu^\mp)$   & \%      & 18.5                            & 14.3                            & 14.5                            \\
		\midrule
		BR$(\tilde{\chi}^\pm_1 \to W^\pm \, \tilde{\chi}^0_1)$    & \%      & 99.4                            & 99.5                            & 99.5                            \\		
		BR$(\tilde{\chi}^\pm_1 \to \tilde{\mu}^\pm_R \, \nu_\mu)$ & \%      & 0.6                             & 0.5                             & 0.5                             \\		
		\midrule		
		$\Delta m(\tilde{\chi}^\pm_1,\tilde{\mu}_R)$              & GeV     & 168.5                           & 207.0                           & 203.4                           \\
		$\Delta m(\tilde{\chi}^0_2,\tilde{\mu}_R)$                & GeV     & 168.0                           & 206.5                           & 202.7                           \\
		$\Delta m(\tilde{\mu}_R,\tilde{\chi}^0_1)$                & GeV     & 7.2                             & 7.8                             & 7.5                             \\
		\bottomrule
	\end{tabular}}
	\caption{\texttt{CheckMATE} analysis results for the benchmarks of table~\ref{tab:benchmark_tab-fully-nonuniv}
	with full gaugino non-universality $M_1<M_2\ll M_3$.}
	\label{tab:rvalBP-fully-nonuniv}
\end{table}
%

\section{Conclusions}
\label{sec:conclusions}

In this paper, we have argued that in order to account for the muon anomalous magnetic moment $g-2$, dark matter and LHC data, non-universal gaugino masses with $M_{1} \simeq 250~{\rm GeV} < M_{2} \ll M_3$ at the high scale are required in the framework of the MSSM.
We also require a right-handed smuon $\tilde\mu_R$ with a mass around 100 GeV
with a small mass gap to neutralino $\tilde{\chi}^0_1$ to evade LHC searches.
The bino-dominated neutralino is a good dark matter candidate due to 
the presence of the nearby right-handed smuon with which it can efficiently co-annihilate in the early universe.
However, the direct detection limits provided by \textsc{XENON1T} provide a strong constraint on this scenario.

We have discussed such a scenario in the framework of an $SU(5)$ GUT
combined with $A_4$ family symmetry, where the three $\overline{5}$ representations form a single triplet of $A_4$ with a unified soft mass $m_F$, while the three $10$ representations are singlets of $A_4$ with independent soft masses
$m_{T1}, m_{T2}, m_{T3}$.
Although $m_{T2}$ (and hence $\tilde\mu_R$) may be light, the muon $g-2$ also requires $M_{1} \simeq 250~{\rm GeV}$ which 
we have shown to be incompatible with 
universal gaugino masses at the GUT scale due to LHC constraints on $M_2$ and $M_3$ arising from gaugino searches.
Therefore, we have allowed non-universal gaugino masses at the GUT scale, which is theoretically allowed 
in $SU(5)$ with non-singlet F-terms.
One should stress that this model is representative of a larger class of such non-universal MSSM scenarios, which can give non-universal masses to left- and right-handed sfermions and which in particular allow a light right-handed smuon with mass around 100 GeV.
After showing that universal gaugino masses $M_{1/2}$ at the GUT scale are excluded by gluino searches, 
we have provided a series of benchmarks which demonstrate that while 
$M_{1}= M_{2} \ll M_3$ is also excluded by chargino searches, $M_{1}< M_{2} \ll M_3$ is currently allowed.
However, there is an unavoidable prediction of our scenario, namely that 
the muon $g-2$ also requires a higgsino mass $\mu \approx -300$ GeV, which --- although consistent with current LHC searches for such higgsino dominated charginos and neutralinos --- will be a target for future such searches.
Although the wino dominated charginos and neutralinos are expected to be somewhat heavier 
and the rest of the SUSY spectrum may have multi-TeV masses outside the reach of the LHC,
the higgsinos with mass of about 300 GeV cannot escape LHC searches,
since they may be pair produced and decay to yield 
muon-dominated di- and tri- lepton plus missing transverse momentum signatures, 
which will be fully probed by the planned increase of total integrated luminosity of up to 3 ab$^{-1}$.
Moreover, the increase of sensitivity of the DM direct detection experiments
by a factor of two, which is expected to take place in the next few years,
will independently probe the entire parameter space of the scenario under study.

To conclude, if the muon $g-2$ turns out to be a true signal of new physics,
then in our scenario we expect a right-handed smuon with mass around 100 GeV, with bino dominated neutralino DM a few GeV lighter,
and a higgsino mass $\mu \approx -300$ GeV.
The {\it whole} such region of MSSM parameter space could be effectively probed in the near future 
and either discovered or excluded by the combined LHC, relic density and DM direct detection experiments as we have discussed above.



\acknowledgments
The authors acknowledge the use of the IRIDIS High Performance Computing Facility, and
associated support services at the University of Southampton, in the completion of this work.
ASB, SFK and PBS acknowledge partial support from the InvisiblesPlus RISE from the European
Union Horizon 2020 research and innovation programme under the Marie Sklodowska-Curie grant
agreement No 690575. SFK acknowledges partial support from the Elusives ITN from the European
Union Horizon 2020 research and innovation programme under the Marie Sklodowska-Curie grant
agreement No 674896.
AB and SFK acknowledges partial  support from the STFC grant ST/L000296/1.
AB also thanks the NExT Institute, Royal Society Leverhulme Trust Senior Research Fellowship LT140094, Royal Society Internationl Exchange grant IE150682 and
Soton-FAPESP grant.
AB also acknowledge the support of IBS centre in Daejeon for the hospitality and support.

\bibliographystyle{JHEP}
\bibliography{bib}

\end{document}